\documentstyle[psfig,conf_iap,]{article}
\begin{document}
\heading{%
%
Sunyaev-Zel'dovich cluster surveys using a large beam
%
} 
\par\medskip\noindent
\author{%
Richard A. Battye$^{1}$, Jochen Weller$^{2}$
}
\address{Jodrell Bank Observatory, Macclesfield, Cheshire SK11 9DL, U.K.}
\address{Institute of Astronomy, Madingley Road, Cambridge CB3 OHA, U.K.}

\begin{abstract}
Sunyaev Zel'dovich cluster surveys can be used to constrain cosmological parameters. Extracting clusters from the primary anisotropies and the unresolved background from very faint clusters is simple when the telescope beam size is small $(\sim 1^{\prime})$, but could be difficult if the beam is larger $(\sim 8^{\prime})$. By reference to examples, we show that this is possible by carefully designing the depth of the survey.
\end{abstract}

\section{Introduction}

The scattering of cosmic microwave background (CMB) photons by hot electrons along the line of sight is known as the Sunyaev-Zel'dovich (SZ) effect~\cite{SZ,BURK}. This is particularly important when the line of sight is crossed by a cluster of galaxies since the temperature of the  intra-cluster medium can reach upto $10^{7}{\rm K}$. Since the optical depth of these clusters is around 0.01, it leads to a distortion of the CMB spectrum which is  a decrement  in the Rayleigh-Jeans region. 

Recently, it has been realized that this effect can be used as a method for reliably selecting clusters. It has some advantages over other methods for cluster selection such as X-rays since the size of the SZ effect for a given cluster is just proportional to the integrated gas pressure along the line of sight, and a number of surveys have been proposed to take advantage of this. One of the most exciting possibilities is that these surveys will be close to mass limited and that it might be possible to use them to provide information on cosmological parameters~\cite{COSMO}, notably the dark energy thought to be driving the accelerated expansion of the universe~\cite{HHM,WBK,BW1}.

One of the key features of any given survey is the integration time on a given field. This should be chosen so as to avoid various sources of confusion which might make it impossible to extract the signal of an individual cluster. This is particularly important when the beam size is large, which in this context means greater than around $4^{\prime}$. In this short exposition of more detailed work presented in ref.~\cite{BW2} we attempt to focus on issues  specific to instruments which have a large beam size, such as the Very Small Array (VSA)~\cite{VSA} which may be upgraded to have sufficient resolution $(\sim 8^{\prime})$ and sensitivity to make an SZ survey worthwhile. 

We will use the model for the cluster distribution described in detail in ref.~\cite{BW2}. Briefly, we use a flat $\Lambda$CDM model with matter density $\Omega_{\rm m}=0.3$, $H_0=72\,{\rm km\,sec}^{-1}{\rm Mpc}^{-1}$ and rms power spectrum amplitude $\sigma_8=0.93$. The cluster mass function is one which fits a series of N-body simulations~\cite{JENK} and we will assume that the individual clusters are virialized and spherical. The hot gas is assumed to have an  isothermal $\beta$-model profile with $\beta=2/3$ and a ratio of the core radius to virial radius of 10. 

We assume that the clusters are observed using a telescope with a Gaussian beam of width $\theta_{\rm FWHM}$ allowing for smoothing to larger beam sizes as described in ref.~\cite{BW2}. Surveys can be parameterized by the flux limit  per beam $S_{\rm lim}$ which we will take to be four times the thermal noise, the angular coverage $\Delta\Omega$ and the frequency $\nu$ of the receivers. For the purposes of this discussion we will use $\nu=30{\rm GHz}$. The results we present here can be converted to other survey frequencies by assuming that the temperature limit per beam is constant.

\section{Sources of confusion}

\subsection{Primary anisotropies}

The primary anisotropies of the CMB are thought to be exponentially suppressed on scales corresponding to multipoles $\ell>2000$ by the effects of photon diffusion. This will create no problem for $\theta_{\rm FWHM}\sim 1^{\prime}$ but can be a problem for instruments with larger beam size. Since the primary motivation for such instruments is usually measuring the angular power spectrum, one has to be careful to avoid it in a cluster survey without eradicating it completely. 

One simple way to do this is to perform a shallow survey covering a large area of sky. Then the thermal noise per beam is large compared with the primary anisotropies. Paradoxically this improves the overall sensitivity of the instrument to the angular power spectrum by reducing the sample variance effects at low $\ell$. Of course, the real features of the primordial anisotropies will be hidden in the map which will have relatively low signal to noise and the measurement of the angular power spectrum will be statistical.

For an interferometric instrument with an $8^{\prime}$ beam, the beam resolution corresponds to the around $\ell\sim 2500-3000$\footnote{The conversion between beam size and angular multipole is different for a single dish experiment.} where the rms of the primary anisotropies is around $3-5\mu{\rm K}$ for currently favoured models. This corresponds to a flux of between 0.6 and 1.0mJy per beam. Clearly, setting $S_{\rm lim}> 5{\rm mJy}$ will alleviate this potential  problem.

\subsection{Unresolved sources}

\begin{figure}
\label{ff}
\centerline{\hbox{
\psfig{figure=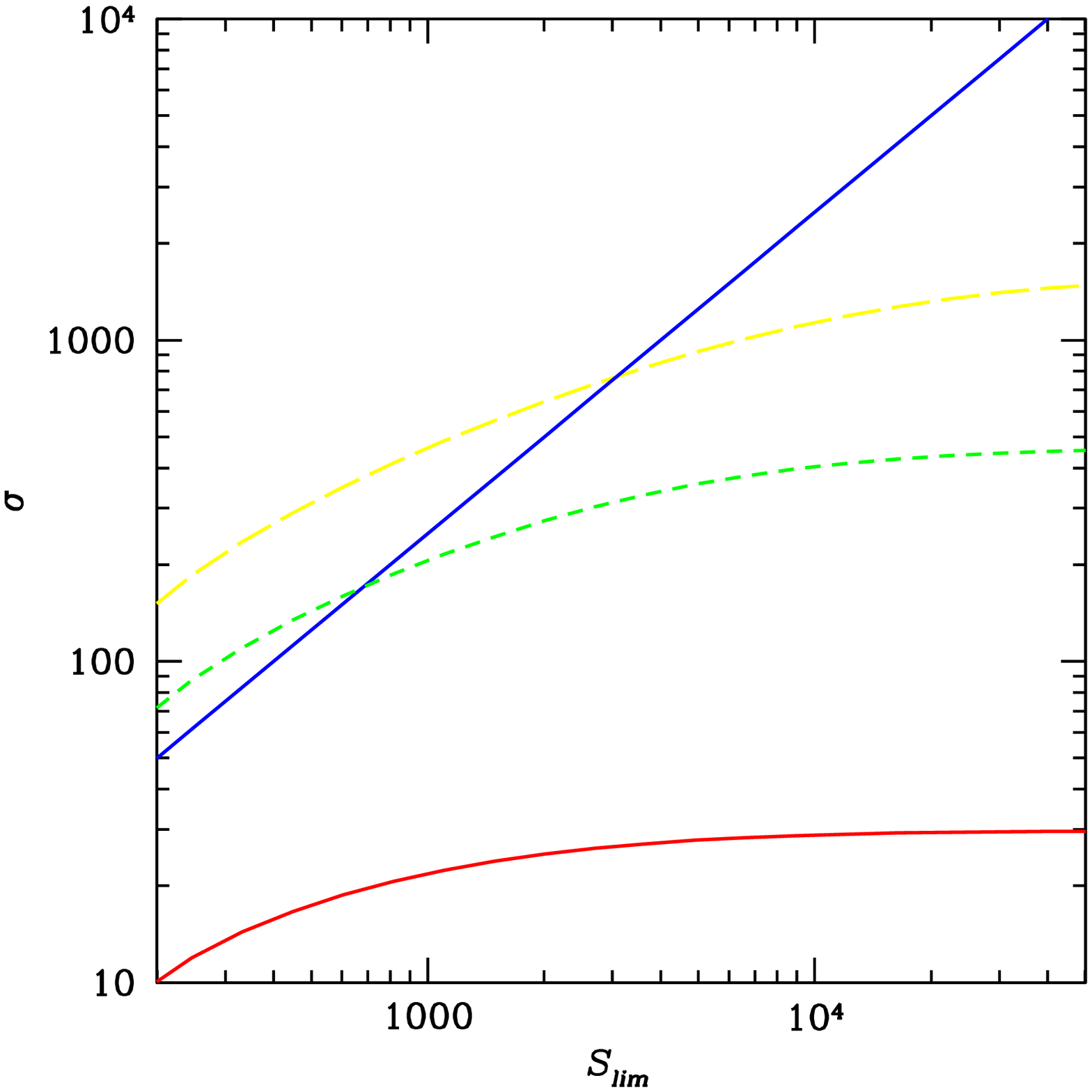,width=6.cm,height=6.cm}
\psfig{figure=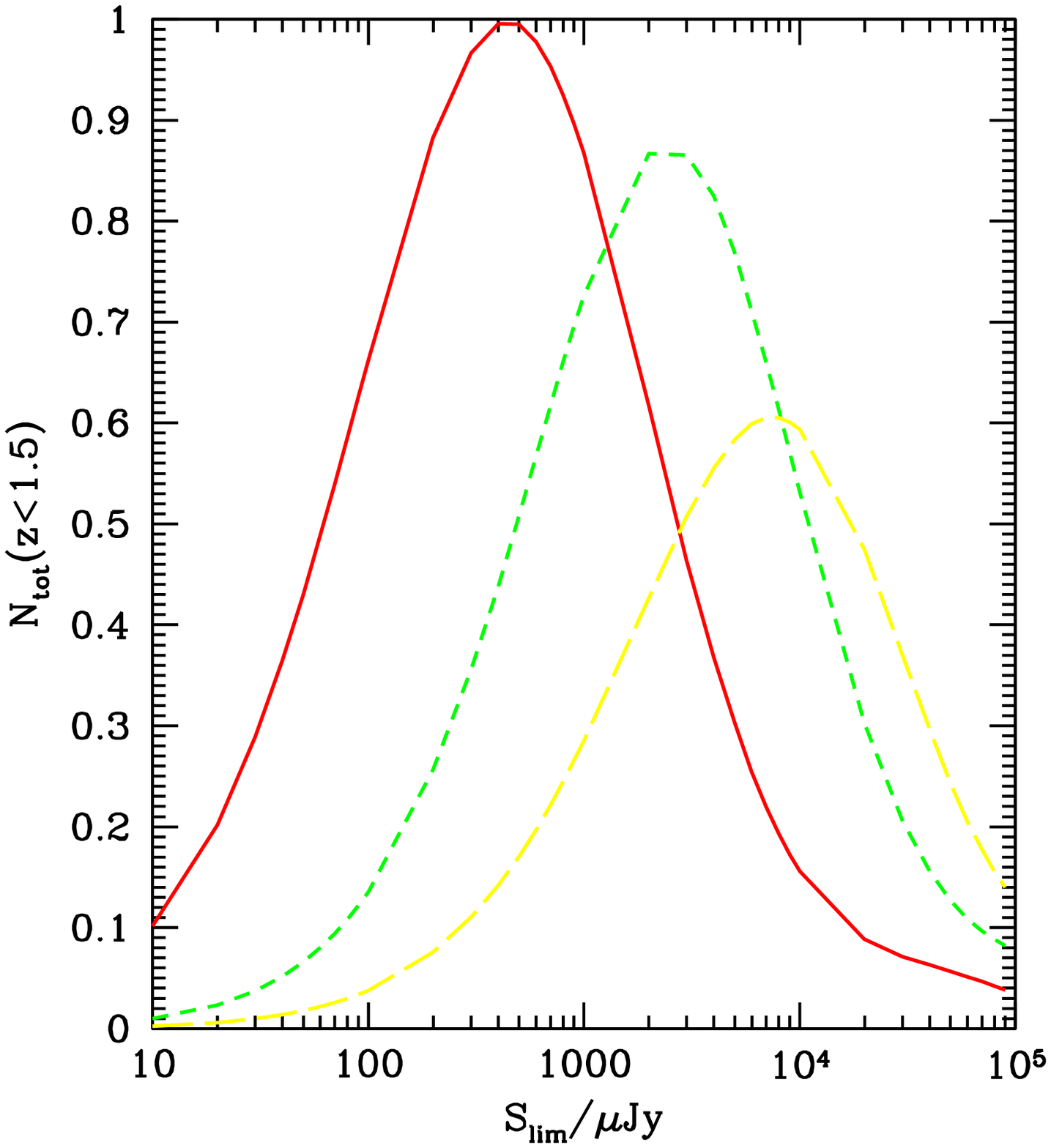,width=6.cm,height=6.cm}
}}
\caption[]{On the left is the residual confusion noise for three different beam sizes ($1^{\prime}$ solid line, $4^{\prime}$ short-dashed line and $8^{\prime}$ long dashed line). The straight line $\sigma=S_{\rm lim}/4$ is also plotted for comparison. On the right is the relative number of clusters which would be found by a survey keeping to the instantaneous sensitivity and total integration time constant.
}
\end{figure}

While the angular power spectrum of the primary anisotropies is decreasing as a function of multipole for $\ell>2000$, that due to cluster population is increasing. If one is trying to select clusters with flux $>S_{\rm lim}$ one should consider combined effect of all of the other unresolved clusters with flux $<S_{\rm lim}$ which create another effective noise source,
\begin{equation}
\sigma=\Omega_{\rm beam}\int_0^{S_{\rm lim}} S^2{dN\over dS} dS\,,
\end{equation}
where $\Omega_{\rm beam}$ is the beam area $(\propto \theta_{\rm FWHM}^2)$ and $dN/dS$ is the differential number count. Clearly, we want this to be less than the thermal noise, $\sigma<S_{\rm lim}/4$.

We have computed this for three curves with $\theta_{\rm FWHM}=1^{\prime},4^{\prime}$ and $8^{\prime}$ respectively and the results are presented in fig.~\ref{ff} (left) along with the line $\sigma=S_{\rm lim}/4$. This shows that a survey with $\theta_{\rm FWHM}=1^{\prime}$ is never confused for sensible values of $S_{\rm lim}$ whereas if $\theta_{\rm FWHM}=4^{\prime}$ or $8^{\prime}$, one will be  confused if $S_{\rm lim}<800\mu{\rm Jy}$ and $4{\rm mJy}$ respectively.

\section{Optimum strategies}

In the previous section we have shown that if the flux limit of the survey is set sufficiently high, then confusion from the primary CMB and unresolved clusters can be avoided even when the beam is large. In the section we attempt to use our assumed knowledge of the cluster distribution to suggest an observational strategy which yields the most clusters. 

In any survey there are two competing effects. Firstly, if one assumes that the clusters are point-like in the beam then at least for nearby clusters the source counts are Euclidean, that is, $N(>S)\propto S^{-3/2}$. Taken in conjunction with the fact that for any given survey specified by the instantaneous sensitivity of the receivers, $\Theta_{\rm FWHM}$ and $\nu$, $S_{\rm lim}\propto \sqrt{\Delta\Omega}$, one might conclude that a very wide survey would find many more clusters. However, in such a wide, and necessarily shallow survey, most of the clusters will be very nearby $(z<0.1)$ and very massive $(M>10^{15}M_{\odot})$. Such objects will be extended in the beam and could be missed if the beam size is small.

We have computed $S_{\rm lim}^2N(>S_{\rm lim},1{\rm deg^2})$ which upto a constant gives the number of clusters with flux $>S_{\rm lim}$ for the family of surveys with $S_{\rm lim}\propto\sqrt{\Delta\Omega}$. We have normalized the curves so as to keep the instantaneous sensitivity of the receivers and the total integration time constant. Then we divided by the largest number of clusters that one could find for any survey. In each case there is an optimum whose flux density increases as $\theta_{\rm FHWM}$ increases. The optima are $\sim 500\mu{\rm Jy}$, $2{\rm mJy}$ and $7{\rm mJy}$ for $\theta_{\rm FWHM}=1^{\prime}, 4^{\prime}$ and $8^{\prime}$ respectively. Furthermore, we see that $\theta_{\rm FWHM}=1^{\prime}$ the survey can be made  optimal, if it is $4^{\prime}$ one can get to within $85\%$ of absolute optimum and even for $8^{\prime}$ one can, by careful selection of observing strategy achieve around $60\%$. For details see ref.~\cite{BW2}.

\section{Conclusion} 

We have show that if the depth of the survey is carefully chosen on avoid potential sources of confusion as well as getting close to the optimum possible even we the beam size is as large as $8^{\prime}$. We suggest that choosing $S_{\rm lim}\sim 7{\rm mJy}$ would be a sensible choice in this respect.

\acknowledgements{We would like to thank R. Kneissl, J. Mohr,  P. Wilkinson, I. Browne, R. Davis and C. Dickinson for numerous helpful comments and insights. RAB is funded by PPARC and JW is funded by the Leverhulme Trust, PPARC and King's College. Some computations were done at the UK National Cosmology Supercomputer Centre funded by PPARC, HEFCE and Silcon Graphics/Cray Research.}

\vfill
\end{document}